\documentstyle[11pt,aaspp4,tighten]{article}

\tightenlines

\def\gsimeq
{\hbox{\raise0.5ex\hbox{$>\lower1.06ex\hbox{$\kern-1.07em{\sim}$}$}}}
\def\lsimeq
{\hbox{\raise0.5ex\hbox{$<\lower1.06ex\hbox{$\kern-1.07em{\sim}$}$}}}
\def\ion#1#2{#1$\;${\small\rm{#2}}\relax}

\begin{document}

\title{An RXTE Observation of NGC 6300: a new bright Compton reflection Dominated
Seyfert 2 Galaxy}

\author{Karen M. Leighly \& Jules P. Halpern}
\affil{Columbia Astrophysics Laboratory, Columbia University, 
550 West 120th Street, New
York, NY 10027, USA, leighly@ulisse.phys.columbia.edu, jules@astro.columbia.edu}
\author{Hisamitsu Awaki}
\affil{Department of Physics, Kyoto University, Kyoto 606-8502, Japan,
awaki@cr.scphys.kyoto-u.ac.jp}
\author{Massimo Cappi}
\affil{Istituto TeSRE/CNR, Via Gobetti 101, 40129 Bologna, Italy,
mcappi@tesre.bo.cnr.it}
\author{Shiro Ueno}
\affil{X-ray Astronomy Group, Department of Physics and Astronomy,
University of Leicester, University Road, Leicester, LE1~7RH, UK, shu@star.le.ac.uk}
\author{Joachim Siebert}
\affil{Max-Planck-Institut \"fur extraterrestrische Physik, Postfach
1603, 85740 Garching, Germany, jos@mpe.mpg.de}

\slugcomment{Accepted for publication in The Astrophysical Journal}


\begin{abstract}

Scanning and pointed {\it RXTE} observations of the nearby Seyfert 2
galaxy NGC~6300 reveal that it is a source of hard X-ray continuum and
large equivalent width Fe K$\alpha$ emission.  These properties are
characteristic of Compton-reflection dominated Seyfert 2 galaxies.
The continuum can be modeled as Compton-reflection; subsolar iron
abundance is required and a high inclination preferred. However, the
poor energy resolution of {\it RXTE} means that this description is
not unique, and the continuum can also be modeled using a ``dual
absorber'', i.e. a sum of absorbed power laws. Observations with
higher energy resolution detectors will cleanly discriminate between
these two models.  Optical observations support the Compton-reflection
dominated interpretation as $L_X/L_{[OIII]}$ is low.  NGC~6300 is
notable because with $F_{2-10}\approx 6.4
\times 10^{-12}\rm\, ergs\,cm^{-2}\,s^{-1}$, it is the second
brightest such object known.
\end{abstract}


\section{Introduction}

The unified model for Seyfert galaxies proposes that orientation of a
molecular torus determines the optical emission-line characteristics
(e.g.\ Antonucci\markcite{1} 1993).  When the molecular torus lies in
our line of sight, it blocks our view of the broad optical emission
lines, leading to a Seyfert 2 classification.  The X-ray emission of
Seyfert 2 galaxies is frequently absorbed (e.g.\ Turner et
al.\markcite{25} 1998; Bassani et al.\markcite{23}
1999), a result which supports this unified model.  In the most
extreme case, the obscuring material presents such a high column
density to the observer that no X-rays are transmitted.  Such objects
are termed ``Compton-thick'' Seyfert 2s, and the only X-rays observed
from them are ones that have been scattered from surrounding material.
(Diffuse thermal X-rays may also contribute).  Such objects are
important because they directly support unified models for Seyfert
galaxies.

The scattering is thought to originate in one or both of two types of
material, each of which imparts characteristic signatures on the
observed X-ray spectrum (for a review, Matt\markcite{15} 1997).
Scattering can occur in the warm optically thin gas that is thought to
produce the polarized broad lines seen in some Seyfert 2s.  The
resulting spectrum has the same slope as the intrinsic spectrum with
superimposed emission lines from recombination. This is the process
which appears to dominate in the archetype Seyfert 2 galaxy NGC~1068
(e.g.\ Netzer \& Turner\markcite{20} 1997).  Scattering can also occur
in optically thick cool material located on the surface of the
molecular torus.  In this case, the process is called Compton
reflection (e.g.\ Lightman \& White\markcite{10} 1988), and the
observed continuum spectrum is flat with superimposed K-shell
fluorescence lines (e.g.\ Reynolds et al.\markcite{22} 1994).
Circinus can be considered the prototype of a Compton-reflection
dominated Seyfert 2 galaxy (Matt et al.\markcite{17} 1996).

Compton-reflection dominated Seyfert 2 galaxies are important because
they may comprise a significant fraction of the X-ray background
(Fabian et al.\markcite{9} 1990).  They were once thought to be rare
(e.g.  Matt\markcite{15} 1997); however, new observations of objects
selected according to their [\ion{O}{III}] emission-line flux, a
method which ideally does not discriminate against highly absorbed
objects, show that they may be more common than previously thought
(Maiolino et al.\markcite{13} 1998).  However, {\it bright} examples
of this class remain rare. This is not surprising, because the
reflected X-rays are very much weaker than the primary continuum.

We present the results of an {\it RXTE} observation of the nearby (18
Mpc) Seyfert 2 galaxy NGC~6300.  This object has a flat hard X-ray
spectrum and huge equivalent width iron line which suggests that it is
a Compton-reflection dominated Seyfert 2 galaxy.  If so, it is one of
the brightest members of this class known, about half as bright as the
prototype, Circinus, and far brighter than other examples.

\section{Data Analysis}

NGC~6300 was first detected in hard X-rays during a {\it Ginga}
maneuver (Awaki\markcite{2} 1991).  We proposed scanning and pointing
observations of this galaxy using {\it RXTE} to confirm the {\it
Ginga} detection.  Another Seyfert 2 galaxy, NGC~6393, was detected
during a {\it Ginga} scan and was also investigated as part of this
proposal.  The data show that NGC~6393 was very faint ($<0.5\rm
\,counts\, s^{-1}$ in the top-layer for 5 PCUs).

The scanning {\it RXTE} observation of NGC 6300 was performed on 1997
February 14--15.  Four of 5 PCUs were on for the entire observation
and analysis was confined to these detectors.  A pointed observation
followed on February 20, 1997, performed with all five PCUs on.  The
data were reduced using Ftools 4.1 and 4.2 and standard data selection
criteria recommended for faint sources.  The resulting exposure for
the pointed observation was 24,896 seconds.  NGC 6300 was detected in
all three layers of the PCA, and the top and mid layers were used for
spectral fitting.  Background subtraction yielded net count rates for
5 PCUs of 4.4 counts s$^{-1}$ (12.5\% of the total) between 3 and 24
keV for the top layer, and 0.86 counts s$^{-1}$ (8.8\% of the total)
between 9 and 24 keV for the mid layer.

The current standard background model for the {\it RXTE} PCA is quite
good; however, NGC~6300 is a relatively faint source and therefore we
attempt to estimate systematic errors associated with the background
subtraction.  Above 30~keV, no signal should be detected; however, we
found a positive signal which could be removed if the background
normalization were increased by 1\%.  Below about 7~keV, no signal
should be detected in the mid or bottom layers. We observed a small
deficit in signal which would be removed if the background
normalization were decreased by 1\%.  We consider this evidence that
the systematic error on the background subtraction is less than 1\%.

The scan observation consisted of 4 passes over the object with a
total scan length of 6 degrees.  The scans were performed keeping the
declination constant during the first two passes and the right
ascension constant during the second two passes.  The resulting scan
profiles when compared with the optical position clearly
indicate that NGC~6300 is the X-ray source.  The field of view of the
PCA is less than two degrees in total width.  Since the scan length
was 6 degrees, and since there are apparently no other X-ray sources
in the field of view, the ends of the scan paths can be used to check
the quality of the background subtraction.  This was of some concern
since NGC~6300 is rather near the Galactic plane ($l=328$, $b=-14$)
and thus there could be Galactic X-ray emission not accounted for in
the background model.  We accumulated spectra with offset from the
source position $>1.5^\circ$.  The exposure time was 1472 seconds.
The count rate between 3 and 24 keV was $-0.28
\pm 0.19 \rm\, counts\,s^{-1}$, so there was no evidence for
unmodeled Galactic emission.  Thermal model residuals show
no pattern; i.e., there is no evidence for a 6.7~keV iron emission
line from the Galactic Ridge (Yamauchi \& Koyama\markcite{28} 1993).

\begin{figure}[t]
\vbox to3.0in{\rule{0pt}{3.0in}}
\includegraphics{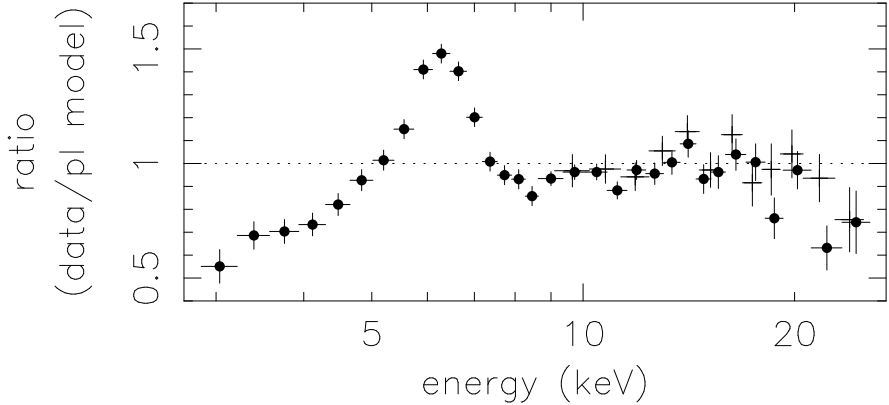}
\caption{Ratio of data to a model consisting of power law plus
Galactic absorption model.  The excess near 6.4~keV clearly indicates
the large equivalent width iron line. Solid points and crosses denote
the top-layer and mid-layer data, respectively.}
\end{figure}

The spectrum from the pointed observation was first modeled using a
power law plus Galactic absorption set equal to $9.38\times
10^{20}\rm\, cm^{-2}$ (Figure 1; Dickey \& Lockman\markcite{6} 1990).
This model did not fit the data well ($\chi^2=609$ for 82 degrees of
freedom (d.o.f.)). The photon index is very flat ($\Gamma=0.60$),
there is clear evidence for an iron emission line and there are
negative low-energy residuals.  Addition of a narrow ($\sigma=0.05\rm
\,keV$) line with energy fixed at $6.4\rm\, keV$ improves the fit
substantially ($\Delta\chi^2=473$) but low energy residuals remain.
Additional absorption in the galaxy rest frame improves the fit
substantially ($\Delta\chi^2=45$).  Freeing the line energy again
improves the fit ($\Delta\chi^2=11$); the best fit rest-frame line
energy is $6.26 \rm\,keV$.  Freeing the line width marginally improves
the fit ($\Delta\chi^2=5$).  The final fit parameters are listed in
Table 1 and fit results are shown in Figure 2.

The absorbed power law plus iron line is an acceptable model.  The
notable properties of the fit are a very flat photon index
($\Gamma=0.68$) and very large equivalent width ($920\,\rm eV$).  Such
parameters suggest that the spectrum of NGC~6300 is dominated by
Compton-reflection (Matt et al.\markcite{17} 1996; Malaguti et
al.\markcite{12} 1998; Reynolds et al.\markcite{22} 1994).  We next use
the {\it pexrav} model in XSPEC to explore this possibility.  This
model calculates the expected X-ray spectrum when a point source of
X-rays is incident on optically thick, predominately neutral (except
hydrogen and helium) material.  The parameter $R$ measures the solid
angle $\Omega$ subtended by the optically thick material:
$R=\Omega/2\pi$.  The model that was fit includes a narrow iron line
and a direct and reflected power law; additional absorption also
appears to be necessary.  The low resolution and limited band pass
provided by the {\it RXTE} spectrum means that not all of the model
parameters could be constrained by the data; thus, the energy of the
exponential cutoff was fixed at 500~keV, approximately the value that
has been found in {\it OSSE} data from Seyfert galaxies (Zdziarski et
al.\markcite{29} 1995), and the inclination was initially fixed
arbitrarily at $45^\circ$.  This model provided a fairly good fit to
the data ($\chi^2=112$ for 77 d.o.f.).  However, the best fit value of
$R$ is very large (1450) and not well constrained.  This indicates
that the spectrum can be modeled using reflection alone, and there is
no significant contribution of direct emission (e.g.\ Matt et al.\markcite{17}
1996).

Reflection alone gives the same $\chi^2$ as the model which includes a
weak direct component, but the fit is still not completely
satisfactory, and it could be improved in two ways.  The first is to
allow the iron abundance to vary.  We set the iron abundance relative
to solar in the {\it pexrav} model equal to the abundances of light
elements and allow these parameters to vary together.  The fit is
improved significantly ($\Delta\chi^2=-36$ for $\Delta$d.o.f.=1) and
gives a slightly subsolar abundance.  The second is to allow the
inclination to be free.  The fit is not very sensitive to this
parameter, as $\Delta\chi^2$ over the whole range is 9.2.  The best
fit value is $\cos(\Theta)=0.22$, corresponding to 77$^\circ$ from the
normal.  The parameters are listed in Table 1.

\begin{deluxetable}{ll}
\small
\tablewidth{20pc}
\tablenum{1}
\tablecaption{Spectral Fitting Results}
\tablehead{
\colhead{Parameter} & \colhead{Value} \\}
\startdata

\multicolumn{2}{c}{Power Law Model} \nl

$\Gamma$ & $0.68_{-0.09,-0.13}^{+0.09,+0.16}$ \nl
$\rm N_H$ ($10^{22}\rm \,cm^{-2}$) &
$5.2_{-1.7,-2.7}^{+1.8,+3.5}$ \nl
$\rm E_{Fe}$ (keV) & $6.26_{-0.06,-0.02}^{+0.06,+0.02}$ \nl
$\rm \sigma_{Fe}$ (keV) & $0.32_{-0.16,+0.07}^{+0.13,-0.07}$ \nl
$\rm F_{Fe}$ ($10^{-5}\,\rm \,cm^{-2}\,s^{-1}$) &
$8.2_{-1.1,+0.5}^{+1.2,-0.4}$ \nl
$\rm EW_{Fe}$ (eV) & $920_{-130,+100}^{+140,-90}$ \nl
$\chi^2$/79 d.o.f. & $75.5_{\ldots,-2.2}^{\ldots,+6.7}$ \nl

\tableline
\multicolumn{2}{c}{Compton Reflection Model} \nl

$\Gamma$ & $1.89_{-0.09,+0.08}^{+0.08,-0.13}$  \nl
Abundance $^a$ & $0.61_{-0.11,+0.15}^{+0.11,-0.16}$  \nl
$\cos(\Theta)$ & $0.22_{-0.22,+0.02}^{+0.16,-0.04}$  \nl
$\rm E_{Fe}$ (keV) & $6.29_{-0.08,+0.02}^{+0.09,-0.02}$ \nl
$\rm \sigma_{Fe}$ (keV) & $0.22_{-0.22,-0.13}^{+0.20,+0.06}$  \nl
$\rm F_{Fe}$ ($10^{-5}\,\rm \,cm^{-2}\,s^{-1}$) &
$4.7_{-1.0,-1.2}^{+1.2,+1.2}$  \nl
$\rm EW_{Fe}$ (eV) & $470_{-100,-150}^{+120,+200}$ \nl
$\chi^2$/77 d.o.f. & $70.3_{\ldots,+9.2}^{\ldots,-1.6}$ \nl

\tableline
\multicolumn{2}{c}{Dual Absorber Model} \nl

$\Gamma$ & $1.71^{+0.21,+0.27}_{-0.19,-0.21}$ \nl
$\rm N_{H(thin)}$  ($10^{22}\rm\,cm^{-2}$) &
$7.7_{-5.2,-1.4}^{+4.0,+2.0}$  \nl
$\rm N_{H(thick)}$ ($10^{22}\rm\,cm^{-2}$) &
$58_{-22,-2}^{+23,+5}$  \nl
$\rm A_{thick}/A_{thin}$$^b$ & $1.9_{-0.9,-0.7}^{+1.4,+1.0}$ \nl
$\rm E_{Fe}$ (keV) & $6.26_{-0.12,+0.002}^{+0.09,-0.005}$ \nl
$\rm \sigma_{Fe}$ (keV) & $0.23_{-0.23,+0.04}^{+0.25,-0.05}$  \nl
$\rm F_{Fe}$ ($10^{-5}\,\rm \,cm^{-2}\,s^{-1}$) &
$3.2_{-1.5,+0.8}^{+1.4,-0.7}$ \nl
$\rm EW_{Fe}$ (eV) & $470_{-220,+50}^{+210,-60}$ \nl
$\chi^2$/79 d.o.f. & $67.6_{\ldots,-1.2}^{\ldots,+2.4}$  \nl
\enddata
\tablecomments{Two kinds of errors are given for each parameter value.  The
first one is the statistical error which represents 90\% confidence
for one parameter of interest ($\Delta\chi^2=2.71$).  The second one
is an estimate of the systematic error obtained by changing the
normalization of the background.  The results of background
under- and oversubtraction by 1\% are given in the sub- and
superscript,  respectively.}
\tablenotetext{a}{Fraction of solar abundance in iron and light
elements.}
\tablenotetext{b}{Ratio of power-law normalizations.}
\end{deluxetable}

We investigated the choice of fixed parameters in the {\it pexrav}
model {\it a posteriori}. Increasing the cutoff energy did not change
the fit.  Decreasing the cutoff to 100~keV produced small differences
in the parameters; namely, the photon index was smaller, the abundance
was higher, and the line flux was lower, but the differences are
within the statistical errors of the adopted model.  We also
investigated the situation when the iron abundance was allowed to
vary, but the abundances of lighter metals were maintained at the
solar value.  A significantly larger photon index by $\Delta\Gamma
\approx 0.15$ was required, due to the decreased reflectivity in soft
X-rays.  We also investigated the effect of a 1\% systematic error in
the background normalization, and the results are listed in Table 1.
This resulted in the largest change in the fit parameters, but the
resulting estimated systematic errors are in the worst case less than
a factor of two larger than the statistical errors.

Because of the low energy resolution of the {\it RXTE} spectra, other
models can be found which fit equally well. It is possible to describe
the spectra using a sum of absorbed power laws (the ``dual absorber''
model; e.g.\ Weaver et al.\markcite{27} 1994).  Specifically, the
model consisted of two power laws, both absorbed by a moderate column,
and one absorbed by a heavy column.  The resulting photon index was
very flat ($\Gamma=1.15$), and is therefore deemed unphysical.
However, including reflection with $R=1$ in the dual absorber model
gave an insignificant improvement in fit ($\Delta\chi^2=-0.6$) but a
more plausible photon index ($\Gamma=1.7$).  The fit parameters are
given in Table 1.  It is notable that the spectra cannot be described
using a highly absorbed transmitted component and an unabsorbed
Compton-reflection dominated component, as has been found to be
appropriate for Mrk~3 (Cappi et al.\markcite{4} 1999).

\begin{figure}[t]
\vbox to4.0in{\rule{0pt}{4.0in}}
\includegraphics{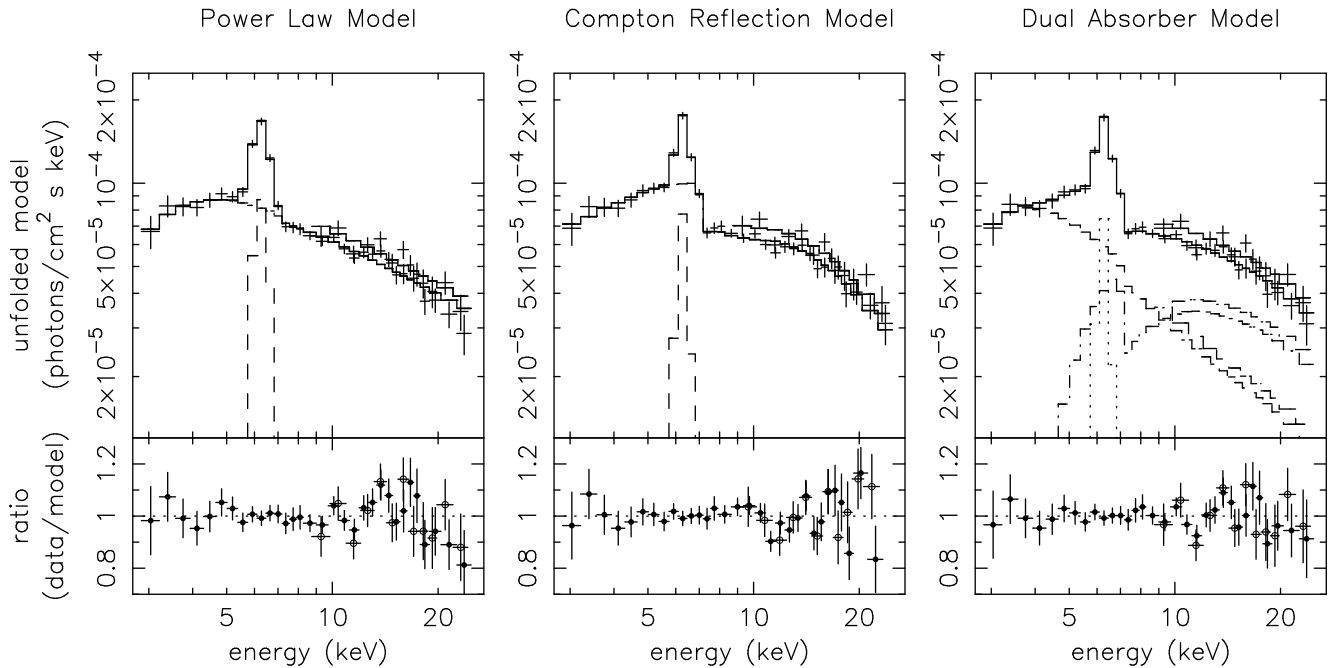}
\caption{Unfolded spectra and ratio of data to model for the three
models considered (see Table 1).}
\end{figure}

\section{Discussion}

\subsection{Compton Reflection Continuum Model}

The X-ray continuum of NGC~6300 can be modeled as pure Compton
reflection.  For solar abundance, the iron line equivalent width
relative to the reflection continuum is predicted to be between 1 and
2 keV depending on inclination (Matt, Perola \& Piro\markcite{18}
1991).  When we fit a power law continuum to the spectra, the
observed equivalent width is nearly 1~keV.  However, when the
reflection continuum is fitted, the measured equivalent width is
reduced to 470~eV.  The reason for the reduction in the measured
equivalent width is that the reflection continuum model includes a
substantial iron edge.  In low resolution data, the iron
line and iron edge overlap in the response-convolved spectra.
Therefore, when the continuum is modeled by a power law, the iron line
models both the line and the edge, so the measured equivalent width is
larger than when the continuum is modeled by the reflection continuum
which includes the iron edge explicitly.  This effect can be seen in
Figure 2.

A significant improvement is obtained when the iron abundance in the
reflection continuum model is allowed to be subsolar.  The iron
abundance is determined by the depth of the iron edge.  Therefore, the
subsolar abundance fits because the iron edge is apparently not as
deep as the model predicts.  The fact that the iron line has a lower
equivalent width than predicted in the Compton reflection continuum
model may also support subsolar abundance.  Alternatively, however,
the apparent subsolar abundances may at least partially be due to
calibration uncertainties in the {\it RXTE} PCA.  (The resolution of
the {\it RXTE} PCA is under some debate; see, Weaver, Krolik \&
Pier\markcite{26} 1998).  The iron line and edge overlap in
the response-convolved spectra.  If the true energy resolution is worse
than the current estimated value, then because the line is an excess
and the edge is a deficit, both would be measured to be smaller than
they really are.

Another source of uncertainty may come from the models themselves,
which depend strongly on the geometry of the illuminating and
reprocessing material.  Models generally assume a point source of
X-rays located above a disk and illuminating it with high covering
fraction.  Such an ideal case may not be attained in nature.

\subsection{Dual Absorber Continuum Model}

The dual absorber model can also describe the spectra.  That
such a fit is successful is not surprising, as any flat continuum can
be modeled as a sum of absorbed power laws (e.g.\ the X-ray
background).  

The iron line equivalent width for the dual absorber model is similar
to that found for the Compton-reflection model, and smaller than that
found for the power law model.  The reason for the difference is that,
like the Compton reflection continuum model, the dual absorber model
explicitly includes an iron edge.  A plausible origin for the iron
line in the dual absorber model is in the absorbing material itself.
However, the iron line equivalent width appears to be too large to
have been produced in the absorbing material.  We investigate this
possibility by comparing the observed iron line flux to the predicted
value from a spherical shell of gas surrounding an isotropically
illuminating point source (Leahy \& Creighton\markcite{11} 1993).  A
line flux of $2.3\times 10^{-5}\rm \, photons\,cm^{-2}\,s^{-1}$ is
predicted for the absorption columns and covering fractions determined
by the fit; this is about half of what is observed ($4.7\times
10^{-5}\rm \, photons\,cm^{-2}\,s^{-1}$).  The dual absorber model
also requires a reflection component with $R=1$ and therefore an
additional iron line with equivalent width $\sim 100$~eV is expected
from the reflection.  Then the predicted flux increases to $3.3\times
10^{-5}\rm \, photons\,cm^{-2}\,s^{-1}$, about 70\% of what is
observed.  The predicted line flux would be smaller if the absorbing
material does not completely cover the source, a circumstance that
would exacerbate the difference between predicted and observed flux.
Thus, the iron line equivalent width, at least to first
approximation, appears to be too large to have been produced in the
absorbing material required by the dual absorber model. Therefore,
either an iron overabundance is required in the dual absorber model,
or the alternative model, the Compton-reflection dominated model, is
favored.

Discrimination between models will come with observations using
detectors with better energy resolution.  If NGC~6300 is a
Compton-thick Seyfert 2, then we should detect soft X-ray emission
lines (e.g.\ Reynolds et al.\markcite{22} 1994).  The lack of any
observable soft excess in the {\it RXTE} data may imply that there is
absorption by the host galaxy (see below), or that there is little
contamination from thermal X-rays or scattering from warm gas.  If the
latter case is true, NGC~6300 will be a particularly clean example of
a Compton reflection-dominated Seyfert 2 galaxy, and the observed soft
X-ray emission lines should be unambiguously attributable to
fluorescence.  Such proof should be easily attainable as NGC~6300 is
bright compared with known Compton-reflection dominated Seyfert 2
galaxies.

\subsection{Information from Other Wavebands}

Optical observations provide some support for Compton-thick absorption
in NGC 6300.  Intrinsically, both hard X-rays and forbidden optical
emission lines should be emitted approximately isotropically;
therefore, the ratio of these quantities should be the same from
object to object.  However, if the absorption is Compton-thick, the
observed hard X-ray luminosity and therefore $L_X/L_{[OIII]}$ will be
significantly reduced; the ratio of the power law to reflection
component 2--10 keV fluxes in the {\it pexrav} model is $\approx 15$.
Care must be taken when applying this test, as the [\ion{O}{III}] flux
must be corrected for reddening in the narrow-line region, and
determination of the reddening using narrow-line Balmer decrements can
be difficult.  The optical spectrum of NGC~6300 is dominated by
starlight (e.g.\ Storchi-Bergmann \& Pastoriza\markcite{24} 1989); to
remove the Balmer absorption from the stars, an accurate galaxy
spectrum subtraction must be done.  Another complication could be
narrow Balmer lines from star formation.  NGC~6300 has a well-studied
starburst ring (e.g.\ Buta\markcite{3} 1987), but H$\alpha$ images
show that the \ion{H}{II} regions are located $>0.5^\prime$
from the nucleus, with a little diffuse emission inside
of that (e.g.\ Evans et al.\markcite{7} 1996).  High quality long-slit
spectra yield $A_v \approx 2.5-3$ from both the red continuum and the
Balmer decrement (Storchi-Bergmann 1999, P.\ comm.).  The observed
2--10 keV flux is $6.4\times 10^{-12}\rm
\,erg\,cm^{-2}\,s^{-1}$ (corresponding to a luminosity of $2.5\times
10^{41}\,\rm erg\,s^{-1}$).  Then $L_X/L_{[OIII]}$ is approximately
1.1--1.9.  This value is quite low and comparable to those obtained from
Compton-thick Seyfert 2s by Maiolino et al.\markcite{13} (1998).  In
particular, Circinus has $F_X=1.4\times 10^{-11} \rm
ergs\,cm^{-2}\,s^{-1}$ (Matt et al.\markcite{16} 1999) and
reddening-corrected $F_{[OIII]}=1.95\times 10^{-11}$ ($A_v=5.2 \pm
0.4$; Oliva et al.\markcite{21} 1994), yielding $L_X/L_{[OIII]}=0.7$.
For comparison, a reddening-corrected sample of Seyfert 1s taken from
Mulchaey et al.\markcite{19} 1994 as a mean ratio of 14.8 ($1\sigma$ range
6.4--33.9).  In contrast, the intrinsic luminosity (i.e.\ corrected
for absorption) for the dual absorber model is $6.9\times 10^{41}\,\rm
erg\,s^{-1}$, implying $L_X/L_{[OIII]} \approx $3.0--5.2.

It is notable also that NGC~6300 shows the reddest continuum toward
the nucleus in a sample of objects studied with long slit spectroscopy
(Cid Fernandes, Storchi-Bergmann \& Schmitt\markcite{5} 1998).
Furthermore, there seems to be a correlation between the presence of a
bar in the host galaxy and presence of a Compton thick Seyfert 2
nucleus (Maiolino, Risaliti \& Salvati\markcite{14} 1999); NGC 6300
has a bar (e.g.\ Buta\markcite{3} 1987).

\subsection{Consistency with Einstein IPC Observation}

The X-ray spectra from Seyfert 2 galaxies frequently includes a soft
spectral component which may originate in scattering by warm gas or
diffuse thermal X-rays.  However, lack of detection in an {\it
Einstein} IPC observation, combined with the {\it RXTE} observation
presented here, shows that there is no evidence for such a component
in NGC~6300.

In 1979 NGC~6300 was observed with the {\it Einstein} IPC for 990
seconds.  It was not detected and the three sigma upper limit to the
count rate was $1.19\times 10^{-2}\rm\, counts\,s^{-1}$ between 0.2
and 4.0 keV (Fabbiano, Kim, \& Trinchieri\markcite{8} 1992). The
Compton reflection-dominated model predicts a count rate of $1.3\times
10^{-2}\rm\, counts\,s^{-1}$, just the same order as the upper limit,
and probably consistent within the uncertainties of the model and the
relative flux calibrations of the two instruments.

There may be intrinsic absorption in the system, for example, from the
host galaxy.  The observed optical reddening of $A_v=2.5-3.0$
corresponds to an absorption column of 4--5$\times 10^{21}\rm
cm^{-2}$, assuming a standard dust to gas ratio.  Including this
column in the Compton-reflection dominated model leads to a predicted
IPC flux of $1.1\times 10^{-2}\rm\, counts\,s^{-1}$, consistent with
the upper limit.  Because the {\it RXTE} band pass is truncated at
3~keV, it is impossible to estimate with accuracy the intrinsic
absorption.  For the Compton-reflection model, the 90\% upper limit
for one parameter of interest is $1.8\times 10^{22}\rm\,cm^{-2}$.

Alternatively, there may have been variability within the 18
years between the IPC and the {\it RXTE} observation.  A probable site
for the reflection is the inner wall of the molecular torus
which blocks the line of sight to the nucleus.  In unified models for
Seyfert galaxies, this material is located outside of the broad line
region, and can be 1--100 pc from the nucleus.  Short term variability
is not expected; long term variability is possible but requires a long
term trend in flux.
	
\acknowledgements

KML acknowledges useful discussions on the {\it RXTE} background with
Keith Jahoda.  KML gratefully acknowledges support by NAG-4112 ({\it
RXTE}) and NAG5-7971 (LTSA).

\clearpage

\clearpage

\normalsize

\end{document}